\documentclass[pre,aps,twocolumn,showpacs,floatfix]{revtex4}

\usepackage[dvips]{graphicx}
\usepackage{amsmath}
\usepackage{amssymb}

\newcommand{\MFPTb}{\mathrm{MFPT}_{\mathrm{w-b}}}
\newcommand{\MFPTw}{\mathrm{MFPT}_{\mathrm{w-w}}}
\newcommand{\MFPT}{\mathrm{MFPT}}

\newcommand{\Sb}{S_{\mathrm{w-b}}(t)}
\newcommand{\Sw}{S_{\mathrm{w-w}}(t)}

\newcommand{\xw}{x_{\mathrm{w}}}
\newcommand{\xa}{x_{\mathrm{a}}}
\newcommand{\xb}{x_{\mathrm{b}}}

\begin{document}

\title{Escape driven by $\alpha$-stable white noises}

\author{B. Dybiec}
\email{bartek@th.if.uj.edu.pl}
\affiliation{M. Smoluchowski Institute of Physics, and Mark Kac Center for Complex
Systems Research, Jagellonian University, ul. Reymonta 4, 30--059 Krak\'ow, Poland }

\author{E. Gudowska-Nowak}
\email{gudowska@th.if.uj.edu.pl}
\affiliation{M. Smoluchowski Institute of Physics, and Mark Kac Center for Complex
Systems Research, Jagellonian University, ul. Reymonta 4, 30--059 Krak\'ow, Poland }

\author{P. H\"anggi}
\email{hanggi@physik.uni-augsburg.de} \affiliation{ Institute of
Physics, University of Augsburg, 86135 Augsburg, Germany \\ and \\
National University of Singapore, Faculty of Science, Physics
Department, Blk S12, 2 Science Drive 3, 117542 Singapore, Singapore}

\date{\today}

\begin{abstract}
We explore the archetype problem of an escape dynamics occurring in
a symmetric double well potential when the Brownian particle is
driven by {\it white L\'evy noise} in a dynamical regime where
inertial effects can safely be neglected. The behavior of escaping
trajectories from one well to another is investigated by pointing to
the special character that underpins the noise-induced discontinuity
which is caused by the generalized Brownian paths that jump beyond
the barrier location without actually hitting it. This fact implies
that the boundary conditions for the mean first passage time (MFPT)
are no longer determined by the well-known local boundary conditions
that characterize the case with normal diffusion. By numerically
implementing properly the set up boundary conditions, we investigate
the survival probability and the average escape time as a function
of the corresponding L\'evy white noise parameters. Depending on the
value of the skewness $\beta$ of the L\'evy noise, the escape can
either become enhanced or suppressed: a negative asymmetry $\beta$
causes typically a decrease for the escape rate while the rate
itself depicts a non-monotonic behavior as a function of the
stability index $\alpha$ which characterizes the jump length
distribution of L\'evy noise, with a marked discontinuity occurring
at $\alpha=1$. We find that the typical factor of ``two'' that
characterizes for normal diffusion the ratio between the MFPT for
well-bottom-to-well-bottom and well-bottom-to-barrier-top no longer
holds true. For sufficiently high barriers the survival
probabilities assume an exponential behavior. Distinct
non-exponential deviations occur, however, for low barrier heights.

\end{abstract}


\maketitle

%%
%% introduction
%%

\section{Introduction}
The noise driven escape from a deterministically metastable state
is a theme that impacts many phenomena in diverse fields of natural sciences
 \cite{kra,bork, trends, jung}.
In particular, as a notable model for a chemical reaction
 Kramers~\cite{kra} pioneered the problem of an escape of a Brownian particle of
 mass $m$ moving in a potential $V(x)$ with local minima
 corresponding to an initial reactant and a final product state.
 In this scenario, both states are assumed to be separated by a
 barrier located at a position $\xb$. In the spatial-diffusion-limited regime, the
 Kramers rate theory is based on a stochastic dynamics that does not involve inertial effects
 and thus is described by an overdamped Langevin dynamics, reading

\begin{equation}
\frac{dx}{dt}=v=-\frac{1}{\eta}V'(x)+\sqrt{\frac{k_BT}{\eta}}\tilde\zeta(t).
\label{lang}
\end{equation}
Here, $\tilde\zeta(t)$ constitutes a white Gaussian noise process with correlations
$\langle\tilde\zeta(t)\tilde\zeta(s)\rangle=2\delta(t-s)$, representing thermal
fluctuations whose intensity is scaled by the friction $\eta$. The escape problem then concerns
the surmounting of an energetic barrier for stochastic trajectories that predominantly dwell the neighborhood
of separating attractors which in this case are made of two neighboring
potential wells. The imposed quasi-stationarity condition is
realized by assuming special boundary conditions with respect to the time evolution equation
for the probability density.
In the overdamped regime, the evolution equation for the probability density $p(x,t)$
follows the Smoluchowski dynamics
\begin{equation}
\frac{\partial p(x,t)}{\partial t}=-\frac{\partial}{\partial x}J(x,t),
\label{smol0}
\end{equation}
where
\begin{equation}
J(x,t)=-\frac{1}{\eta}V'(x)p(x,t)-\frac{k_BT}{\eta}\frac{\partial
p(x,t)}{\partial x},
\label{smol}
\end{equation}
and the stationarity approximation describes escape events which
correspond to a constant, non-vanishing flux of probability
$J_s=J(x)$. Those stochastic realizations of the process that have
surpassed the barrier top are immediately absorbed and re-inserted
into the original attractor region. In this way a steady probability
flow across the activated barrier state located between the locally
stable states of ``reactants'' and ``products'' is established.
Escaped trajectories, absorbed at a position larger than the barrier
location $\xa>\xb$ require that $p_s(\xa)=0$ on the whole half-line
$\xa > \xb$. The rate formulation is then based on the ``flux over
the population'' method~\cite{bork,flux}, yielding in this case of a
Smoluchowski dynamics the celebrated result
\cite{kra,bork,trends,jung}:

\begin{equation}
k=\frac{\omega_b}{\eta}\frac{\omega_w}{2\pi}\exp\left[-\frac{\Delta V}{k_B T}\right],
\label{rate}
\end{equation}
with $\Delta V$ denoting the activation energy $\Delta
V=V(\xb)-V(\xw)$ and $\omega_w, \omega_b$ representing the
frequencies of the particle's motion near the metastable potential
minimum $\xw$ and at the top of the crossed barrier, respectively.
This result represents a feasible estimate for the actual reaction rate, if
all trajectories ejected by the source properly thermalize before
eventual thermal fluctuations drive them out of the initial well and, most importantly,
a distinct time-scale separation between escape dynamics and intra-well relaxation holds true.
This latter requirement of a clear-cut time-scale separation is at the basis for the description of the
escape dynamics in terms of a (time-independent) rate coefficient \cite{bork,flux}.

Yet another alternative to the approaches discussed above is rooted
in the concept of the mean first passage time (MFPT), i.e. the
average time that a random walker starting out from a point $x_0$
inside the initial domain of attraction, assumes in order to leave
the attracting domain for the first time \cite{bork,trends, jung}.
Put differently, the MFPT is the average time needed to cross the
deterministic separatrix-manifold for the first time \cite{bork, BC,
talk}. At weak noise the MFPT becomes essentially independent of the
starting point, i.e. $t(x_0)\approx t_{\MFPT}$ for all starting
configurations away from the immediate neighborhood of the
separatrix. Given the fact that the crossing of the separatrix in
either direction equals for {\it normal diffusion} one half, the
total escape time $t_e$ equals $2t_{\MFPT}$ and thus the rate of
escape $k$ itself becomes in this case:
\begin{eqnarray}
k=\frac{1}{2t_{\MFPT}} \;.
\end{eqnarray}

Characterization of the escape rate by use of the MFPT is a rather complex notion
for a general class of stochastic processes.
 In particular, the MFPT analysis requires the choice of a correct boundary
 condition \cite{BC, talk}. These are well known for one-dimensional
 stochastic diffusion Markov processes $x(t)$ which are of the Fokker-Planck
 form, Eqs.~(\ref{smol0}--\ref{smol}) or for one dimensional master equations
 with birth and death kinetics. With generally non-Gaussian white noise the
 knowledge of the boundary location alone typically cannot
specify in full the corresponding boundary conditions for, say,
absorption or reflection, respectively
\cite{BC, former}. In particular, the
trajectories driven by non-Gaussian white noise depict
discontinuous jumps. As a consequence, the location of the
boundary itself is not hit by the majority of discontinuous sample
trajectories. This implies that regimes beyond the location of the
boundaries must be properly accounted for when setting up the
boundary conditions. Most importantly, returns (i.e. so termed
re-crossings of the boundary location) from excursions beyond the
specified state space back into this very finite interval where
the process proceeds must be excluded.

Following our reasoning in discussing L\'evy-Brownian motion on
finite intervals \cite{former}, we present in this paper the
analysis of escape events of a non-inertial, generalized diffusion
process which is driven by L\'evy noise dwelling a symmetric double-well potential.\\

\section{White L\'evy noise}

The standard definition of the Gaussian white noise specifies the
latter as a time derivative of the Wiener process (i.e. a
derivative of a stationary process with independent and
Gaussian-distributed increments whose covariance is given by
$\langle W(t)W(s)\rangle=\textrm{min}(t,s)$). A classical Brownian
motion (the Wiener process) can be therefore represented as a
limit in distribution of independent Gaussian jumps taken at
infinitesimally short time intervals of non-random length $1/n$.
Alternatively (following e.g. definition of Feller
\cite{feller,zolotarev,gnedenko}), the Wiener process can be
thought of as a limiting process of random Gaussian jumps at
random Poissonian jump times:

\begin{eqnarray}
W(t) & = & \lim_{n\rightarrow\infty}\sum^{[tn]}_{k=1}W_k\left( \frac{1}{n} \right)
 = \lim_{n\rightarrow\infty}W\left (\frac{N(nt)}{n}\right) \nonumber\\
& = & \lim_{n\rightarrow\infty}\sum^{N(tn)}_{k=1}W_k,
\label{wiener}
\end{eqnarray}

The symbol $[tn]$ stands for an integer number of jumps which for
a Poisson counting process $N(nt), t\geqslant 0$ with them mean
$\langle N(nt)\rangle=nt$ are assumed to be independent
(decoupled) of {\it i.i.d} random variables $W_k$ sampled from the
Gaussian distribution. The equality sign in Eq.~(\ref{wiener})
denotes a limit in distribution sense. For $n\rightarrow \infty$,
the Poisson distribution becomes peaked around $k=nt$ and the
limiting process $W(t), t\geqslant 0$ tends to a Brownian motion
diffusion for which the cumulative distribution function reads
\cite{foot}

\begin{widetext}
\begin{eqnarray}
\mathrm{Prob}\left \{ W(t)\leq w \right \} & = & \lim_{n\rightarrow\infty}\sum^{\infty}_{k=0}
\mathrm{Prob}\left\{ W\left(\frac{k}{n}\leq w \right) \right\} \mathrm{Prob}\left\{ N(nt)=k\right\} \nonumber \\
& = &
\lim_{n\rightarrow\infty}\int^{w}_{-\infty}\sum^{\infty}_{k=0}(2\pi
k/{n})^{-1/2}\exp{(-nx^2/2k)} \frac{(nt)^k}{k!}\exp{(-nt)}dx
=\int^{w}_{-\infty}\frac{e^{-x^2/2t}}{\sqrt{2\pi t}}dx.
\end{eqnarray}
\end{widetext}

Here, we introduce a
non-Gaussian white noise as a derivative of the generalized Wiener
process $W_{\alpha,\beta}(t)$, i.e. a non-Gaussian random process
with stationary and independent increments. The increments
of such generalized Wiener process have the $\alpha$-stable
distribution with the stability index $\alpha$ and the time
increment $\Delta t^{1/\alpha}$ as a scale parameter:

\begin{eqnarray}
W_{\alpha,\beta}(t) & = & \int^t_0\zeta(s) ds=\int^t_0
dL_{\alpha,\beta}(s)\nonumber \\
& \approx & \sum\limits_{i=0}^{N-1}(\Delta s)^{1/\alpha}\zeta_i,
\end{eqnarray}
where $\zeta_i$ are independent random variables distributed with
the stable, L\'evy probability density function (PDF)
$L_{\alpha,\beta}(\zeta;\sigma,\mu=0)$ and $N\Delta s=t-t_0$.
The parameter $\alpha$ denotes the stability index, yielding the
asymptotic power law for the jump length distribution being
proportional to $|\zeta|^{-1-\alpha}$. The parameter $\sigma$
characterizes a scale, $\beta$ defines an asymmetry (skewness) of
the distribution, whereas $\mu$ denotes the location parameter.
Throughout the paper, we deal only with strictly stable
distributions not exhibiting a drift regime; this implies a
vanishing location parameter $\mu=0$ in the remaining part of this
work. For $\alpha\not=1$, the characteristic function $\phi(k) =
\int_{-\infty}^\infty d\zeta e^{-ik\zeta}
L_{\alpha,\beta}(\zeta;\sigma,\mu)$ of an $\alpha$-stable random
variable $\zeta$ can be represented by
\begin{equation}
\phi(k) = \exp\left[ -\sigma^\alpha|k|^\alpha\left(
1-i\beta\mbox{sign}(k) \tan\frac{\pi\alpha}{2} \right)\right],
\label{charakt}
\end{equation}
while for $\alpha=1$ this expression reads
\begin{equation}
\phi(k) = \exp\left[ -\sigma|k|\left(
1+i\beta\frac{2}{\pi}\mbox{sign} (k) \ln|k| \right) \right] \;.
\label{charakt1}
\end{equation}
The three remaining parameters vary within the allowed regimes $
\alpha\in(0,2],\; \beta\in[-1,1],\; \sigma\in(0,\infty)$. Random variables $\zeta$
corresponding to the characteristic functions
(\ref{charakt}) and (\ref{charakt1}) can be generated using the Janicki--Weron algorithm
\cite{weron1996,janicki}.

For $\alpha=2$ (and an arbitrary skewness parameter $\beta$, cf.
Eq.~(\ref{charakt})), the generalized L\'evy-Brownian motion
$W_{\alpha,\beta}(t)$ becomes a standard Gaussian Wiener process
whose time derivative leads to the well known Gaussian white noise
limit.

Following this interpretation, the stochastic source term in
Eq.~(\ref{lang}) can be represented as a sum of independent pulses
(having stable distribution) acting on equally spaced times. As a
consequence, such L\'evy noise is white, i.e. its autocorrelation
function is formally a Dirac-delta function of time.

For the sake of clarification, we mention here that there exist
different representations of other L\'evy noises that are
non-Markovian in nature. As an example, so called fractional
Gaussian (or L\'evy) noise is sometimes defined in literature as
the time derivative of a fractional Brownian motion process
\cite{mandelbrot,chaudhury06,zaslavsky,sokolov,kou}. In contrast to the
Gaussian (or our L\'evy) white noise, the fractional Gaussian
noise (fractional L\'evy stable motion) may exhibit slowly
decaying time-correlations. This is however not the case with the
type of L\'evy noise addressed in our paper, where the noise
source at the level of the Langevin equation {\it is a white noise
process}.

\section{Escape in a double well: Survival probability and mean first passage time }
Let us consider a Brownian, overdamped particle in an external
potential $V(x)$, i.e.,
\begin{equation}
V(x)=-\frac{a}{2}x^2+\frac{b}{4}x^4.
\label{potentialeq}
\end{equation}
which is driven by L\'evy stable white noise. The sample
trajectories are then obtained by a direct integration of
Eq.~(\ref{lang}):
\begin{equation}
x(t) = x_0-\int_{t_0}^{t} V'(x(s)) ds + \int_{t_0}^{t}
dL_{\alpha,\beta}(s),
\label{lcalka}
\end{equation}
using the standard techniques of integration of stochastic
differential equation with respect to the L\'evy stable PDFs
\cite{former,weron1995,weron1996,ditlevsen1999,janicki,newman1999,dybiec2004}.
The first passage time problem has been analyzed as
$\tau=\mathrm{inf}\{ t\geq 0 \;|\; x(t)\geq \xb \}$ with
trajectories $x(t)$ starting at $x=x_0$ and subject to corresponding
boundary conditions. In particular, an absorbing boundary condition
is realized by stopping the trajectory whenever it reaches the
boundary, or, more typically, whenever it has jumped beyond that
very boundary location. The role of reflection, which in the case of
a free diffusion \cite{former} has been assured by wrapping the
hitting (or crossing) trajectory around the boundary location, while
preserving its assigned length, see in Ref. \cite{former,
dybiec2004}, is taken over naturally here by the confining potential
walls of the symmetric double well. The details of our employed
numerical scheme for stochastic differential equations driven by
L\'evy white noise has been detailed elsewhere \cite{former}. In the
following we shall omit cases when $\alpha= 1$ with $\beta\ne0$. In
fact, this parameter set is known to induce instabilities in the
numerical evaluation of corresponding trajectories
\cite{weron1995,weron1996,ditlevsen1999,janicki,dybiec2004}.

\begin{figure}[!ht]
\includegraphics[angle=0, width=8.0cm, height=5cm]{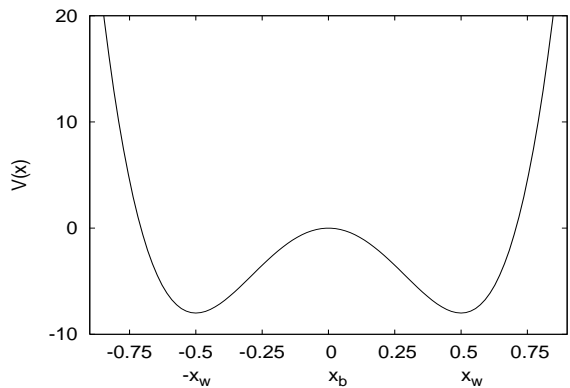}
\caption{The generic double well potential
$V(x)=-\frac{a}{2}x^2+\frac{b}{4}x^4$ for inspection of the
Kramers problem discussed in the paper. To assure sufficiently
high barrier separating the stationary states $-\xw, \xw$, the
following set of parameters has been chosen: $a=128,b=512,\Delta
V=V(0)-V(-\sqrt{a/b})=8$.} \label{potential}
\end{figure}

In the following we consider several differing situations. For a
particle that starts out at a well bottom $x_0=-\xw$, cf. Fig.~\ref{potential} and makes excursions
toward the neighboring well bottom, taken as an absorbing boundary
we use the subscript notation $w-w$. Likewise, fore a particle
staring out at well bottom and being absorbed at barrier top
location we use the notation $w-b$. Using the statistics of the firs
passage time events we shall next study the two corresponding
survival probabilities
\begin{eqnarray}
S_{\mathrm{w-b}}(t) &=& 1 - {\mathcal F_{\mathrm{w-b}}(t)} \\
S_{\mathrm{w-w}}(t) &=& 1 - {\mathcal F_{\mathrm{w-w}}(t)} \;,
\end{eqnarray}
where $ \mathcal F_{...}(t)$ is the corresponding cumulative first
passage time distribution, i.e.,
\begin{equation}
\mathcal F_{\mathrm{w-b}, \mathrm{w-w}}(t)= \int^{\xb,\xw}_{-\infty}
p_{\mathrm{w-b}, \mathrm{w-w}}(x,t) dx \;,
\end{equation}
with $p_{\mathrm{w-b}, \mathrm{w-b}} (t)$ being the corresponding
first passage time density.

In order for the escape to become dominated by a clear-cut, single
time-scale we use a sufficiently high potential barrier, thereby
enforcing rare escape events. A too low barrier would involve too
many recrossing events with the escape then being ruled by many time
scales. Our used parameters for the potential are the parameter set
$a,b$ with $a=128, b=512$, yielding a barrier height of $\Delta
V=V(0)-V(-\sqrt{a/b})=\frac{a^2}{4b}=8$. This symmetric potential
well is schematically depicted with Fig.~\ref{potential}.

\subsection{Case of normal diffusion}

For $\alpha=2$, the process driven by white L\'evy stable noise
approaches the Brownian limit. The $\alpha$-stable white L\'evy
noise, as used in this study, then leads to the standard Gaussian
white noise with intensity $2\sigma^2$, i.e. $\langle \zeta(t)
\zeta(s) \rangle_{\alpha=2} = 2\sigma^2\delta(t-s)$. In this case,
the corresponding values of the MFPT from the potential minimum
$-\xw$ to the top of the absorbing potential barrier, i.e. $\xb$ or
to the neighboring, absorbing minimum, $x=\xw$, can be calculated
from the following quadrature formulas
\begin{eqnarray}
\mathrm{MFPT}(-\xw\to x) & = & \frac{1}{\sigma^2}
\int\limits_{-\xw}^{x}\exp\left[ V(z)/\sigma^2 \right] \nonumber \\
& \times & \int\limits_{-\infty}^{z}\exp\left[ -V(y)/\sigma^2 \right]dydz.
\label{mfpttheory}
\end{eqnarray}
Note that for a system driven by white Gaussian noise the energy
difference $\Delta V$ measured in units of $k_BT$ may be directly
related to the intensity of the noise $\sigma^2$. In contrast, for
L\'evy stable noises with $\alpha<2$, the scale parameter $\sigma$
is no longer ``thermodynamically'' related to the system temperature
and consequently becomes a free parameter of the model. We have set
throughout this study this value to $\sigma=\sqrt{2}$.

\subsection{Survival probabilities for $\alpha$-stable noise driven
escape}

\begin{figure}[!ht]
\includegraphics[angle=0, width=8.0cm, height=5cm]{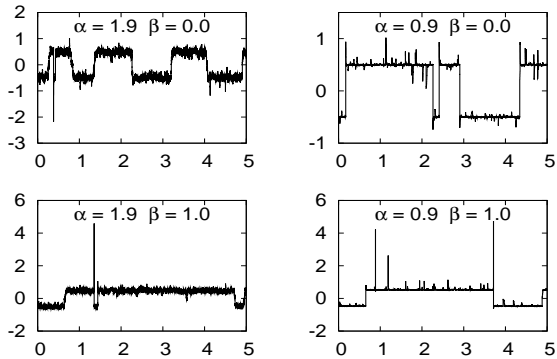}
\caption{Sample trajectories of random walks in the generic double
well potential subjected to L\'evy stable noise with $\alpha=1.9$
(left panel) and $\alpha=0.9$ (right panel) with various $\beta$
($\beta=0$, top panel, $\beta=-1.0$, bottom panel). The scale
parameter is $\sigma=\sqrt{2}$, the chosen time step of integration
is $dt=10^{-4}$.} \label{trajectories}
\end{figure}

Typical sample trajectories of the stochastic process defined by
Eq.~(\ref{lang}) are depicted in Fig.~\ref{trajectories}. We observe
that a non-zero $\beta$ parameter induces a dynamical asymmetry for
the escape dynamics occurring in a {\it symmetric} double well
potential. It is reflected in our numerical simulations by the
emergence of stochastic trajectories that spend more time in the
vicinity of one of the potential minima. As a consequence, one of
the attracting states of the process ($\pm \xw$) becomes
favored over the other. This kind of behavior can be also detected
in our other figures. Notably, for a decreasing value of the
stability index $\alpha$ we observe larger fluctuations of the
particle positions. These occasional long jumps of trajectories may
be of the order of, or even larger than the distance $2
\xw$ separating the two minima of the symmetric potential
$V(x)$.

%
% trajectories
%

\begin{figure}[!ht]
\includegraphics[angle=0, width=8.0cm]{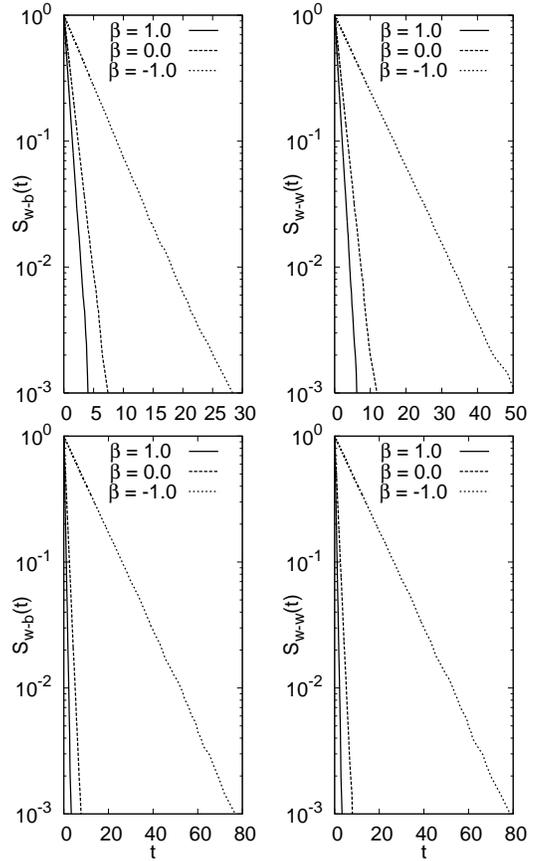}
\caption{Survival probability densities $\Sb$ (left panel) and $\Sw$
(right panel) for a particle dwelling the generic double well
potential subjected to L\'evy stable noise with $\alpha=1.9$ (top
panel) and $\alpha=0.9$ (bottom panel). The simulation parameters
are: $dt=10^{-5}$, $N=2\times 10^4$, $\sigma=\sqrt{2}$.}
\label{survival}
\end{figure}

From the ensemble of single trajectories, which are subject to the
boundary conditions discussed above, and presented in
Fig.~\ref{trajectories}, we estimated the survival probability
densities $\Sb$ and $\Sw$, cf. see Fig.~\ref{survival}.
 The behavior of the
survival probability is consistent with the results as predicted by
inspection of the corresponding stochastic trajectories: We clearly
detect from Fig.~\ref{survival} that at a chosen value of $\alpha$ a
decrease of the skewness $\beta$-parameter of a driving white noise
causes a distinct decrease the rate of escape of the
 particle from the left potential minimum; thus stabilizing the starting position at $x=-x_\mathrm{w}$.
Most importantly, at sufficient high barrier heights a visible
exponential decay of
 the survival probability occurs, -- according to $S(t)=\exp(-t/T_{MFPT})$. This is the familiar behavior for regular Brownian motion
 and has been detected already previously for symmetric L\'evy noises \cite{chechkin2005,imkeller} and for totally skewed (with $\beta=1$)
 one sided L\'evy motions \cite{eliazar}.
Here we observe it as well
 for {\it skewed stable noises}. Notably, the characteristic exponent $T_{MFPT}$ clearly depends on
both noise parameters, i.e. on the value of the stability index
$\alpha$ and also on the skewness parameter $\beta$.

Our numerical analysis indicates that typically the survival
probability $\Sw$ for the ``well-bottom-to-well-bottom'' setup
exceeds the survival probability $\Sb$ for the
``well-bottom-to-barrier-top'' setup. This observation can be
explained by features of a noise-driven dynamics: A particle
performing the motion under the influence of noise needs more time
to dwell the neighboring potential minimum than is needed to reach
the top of the barrier. This kind of behavior can be nevertheless
weakened by diminishing the stability parameter $\alpha$. For
$\alpha<2$ the trajectory of the random particle becomes
discontinuous and at sufficiently small $\alpha$ a particle, on
average, completes its escape from the left potential minimum by a
long jump which can overpass the right potential minimum.
Consequently, both survival curves start to overlap converging to
the same function at a given small stability index $\alpha$.

%
% survival probabilities
%

%
% S(t) for Cauchy
%
\begin{figure}[!ht]
\includegraphics[angle=0, width=8.0cm, height=5cm]{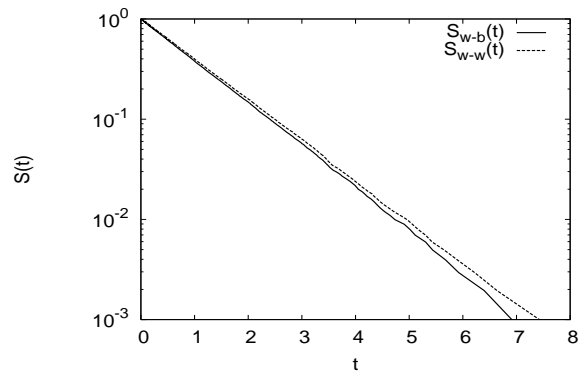}
\caption{Survival probability densities $\Sb$ and $\Sw$ for a
particle wandering in the generic double well potential subjected to
Cauchy noise, i.e. L\'evy stable noise with $\alpha=1.0$.
$\MFPTb=1.038\pm 0.008$ and $\MFPTw=1.086\pm 0.008$. The simulation
parameters are: $dt=10^{-5}$, $N=2\times 10^4$, $\sigma=\sqrt{2}$.}
\label{cauchy}
\end{figure}

For the Cauchy limit, i.e., $\alpha=1$, an approximate result for
the mean crossing time as a function of the noise strength $\sigma$
has been given in Ref. \cite{chechkin2005} by use of the fractional
Fokker-Planck equation. In order to compare results obtained in this
work with the former studies presented by Chechkin {\it et al.}
\cite{chechkin2005}, we have used a re-scaled form of
Eq.~(\ref{lang})
\begin{equation}
\frac{dx}{dt}=(x-x^3)+\zeta(t),
\label{rescaled}
\end{equation}
obtained by a set of transformations $x\to x/x_{{\mathrm w}}$, $t\to
t/\tau$ with $\xw^2=a/b$, $\tau=\eta/a$ and consequently
$\sigma^\alpha=D \to \sqrt{\frac{k_BT}{\eta}}\tau/\xw^\alpha$. In
these units $\sigma\to \frac{2\sigma}{128^{1/\alpha}}$, thus,
$\sigma_{\alpha=2}=\frac{1}{4}$ and
$\sigma_{\alpha=1}=\frac{\sqrt{2}}{64}$. From the approximate
formula derived in \cite{chechkin2005} for a strictly weak noise
case at $\alpha=1$ one obtains $\mathrm{MFPT}\approx 141.71$ ,
whereas our simulations for this noise strength yield
$\mathrm{MFPT}\approx 132.86$, which differs only by ca. 7\%.
Exemplary survival probabilities $\Sb$ and $\Sw$ for this
Cauchy-L\'evy noise are presented in Fig.~\ref{cauchy}. The data
 indicate that for the Cauchy noise the MFPT for
well-bottom-to-barrier-top scenario is somewhat smaller than the
MFPT for the case of well-bottom-to-well-bottom, cf.
Fig.~\ref{cauchy}.

\begin{figure}[!ht]
\includegraphics[angle=0, width=8.0cm, height=5cm]{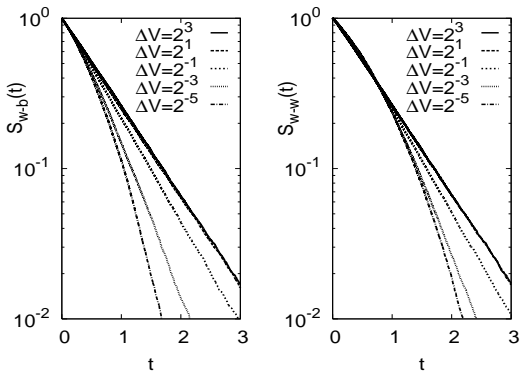}
\caption{Survival probabilities $\Sb$ (left panel) and $\Sw$ (right
panel) for various barrier heights separating two wells of the
potential. The system is driven by L\'evy-Smirnoff noise, i.e.
L\'evy stable noise with $\alpha=0.5$ and $\beta=1$. The simulation
parameters are: $dt=10^{-5}$, $N=2\times 10^4$, $\sigma=\sqrt{2}$.}
\label{height}
\end{figure}

Our prior studies of a {\it free} L\'evy-Brownian motion driven
by $\alpha$ stable noise has shown that for some
parameterizations of the L\'evy noise (totally skewed stable
distributions with $\alpha<1$) a non-exponential survival
probability density emerges. Therefore, in this study we tested
whether for a particle driven by a L\'evy-Smirnoff noise, i.e.
with setting $\alpha=0.5,\;\beta=1$, a decrease of the potential
barrier may induce visible deviations from the exponential
distribution.
 The results of the test are displayed in
Fig.~\ref{height} where the survival probability densities $\Sb$ and
$\Sw$ for various heights of the potential barrier $\Delta V$ are
presented. For decreasing $\Delta V$ the motion of a particle indeed
approaches the behavior of a free diffusion; consequently also
distinct deviations from the exponential character of survival
distributions do show up.

%
% survival probability for various barrier heights
%

\subsection{ Behavior for the MFPT}

\begin{figure}[!ht]
\includegraphics[angle=0, width=8.0cm,height=5cm]{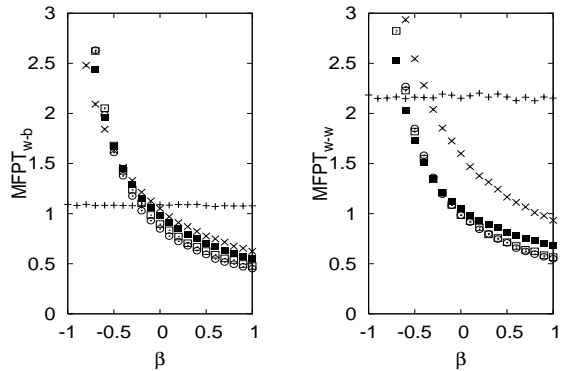}
\caption{Mean first passage times for well-bottom-to-barrier-top
 $\MFPTb$ (left panel) and mean first passage times for well-bottom-to-well-bottom
$\MFPTw$ (right panel) as a function of the skewness parameter $\beta$.
 Simulation parameters $dt=10^{-5}$, $N=2\times 10^4$,
$\sigma=\sqrt{2}$. The error bars estimated using bootstrap method
with $N_b=2\times 10^3$. The white L\'evy stable noise with the
stability index $\alpha=2$ and any allowed value of $\beta$ is
equivalent to the white Gaussian noise, what is manifested by the
independence of MFPT$(\beta)$ for $\alpha=2$. The various symbols
represent the various values of the stability index $\alpha$: `$+$'
$\alpha=2.0$, `$\times$' $\alpha=1.9$, `$\circ$' $\alpha=1.5$,
`$\square$' $\alpha=1.3$, and `$\blacksquare$' $\alpha=1.1$.}
\label{topandwell_2d}
\end{figure}

\begin{figure}[!ht]
\includegraphics[angle=0, width=8.0cm, height=5cm]{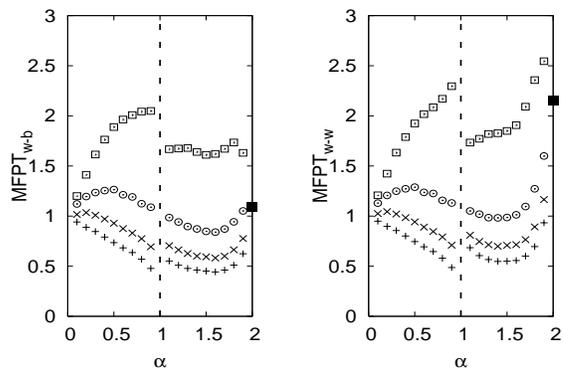}
\caption{Mean first passage times for well-bottom-to-barrier-top $\MFPTb$ (left panel)
and mean first passage times for well-bottom-to-well-bottom $\MFPTw$ (right panel) as a
function of the stability index $\alpha$.
 Simulation parameters $dt=10^{-5}$, $N=2\times 10^4$, $\sigma=\sqrt{2}$ and $N_b=2\times 10^3$.
 The black squares indicate equality of MFPTs for $\alpha=2$ with any $\beta$.
 The various symbols represent the various values of the skewness parameter $\beta$:
`$+$' $\beta=1.0$,
`$\times$' $\beta=0.5$,
`$\circ$' $\beta=0.0$,
and `$\square$' $\beta=-0.5$.
}
\label{atopandwell_2d}
\end{figure}

From the ensembles of collected first passage times we have also
evaluated directly the mean values of the distributions. Our
findings for MFPTs are presented in
Figs.~\ref{topandwell_2d}--\ref{atopandwell_2d}. The depicted
results corroborate with the decrease of the escape rate (i.e. the
inverse of the MFPT) upon increasing the skewness $\beta$. Moreover,
for the stability index $\alpha=2$ (normal Brownian white noise) and
for any skewness parameter $\beta$, the Gaussian case is properly
re-confirmed, cf. Figs.~\ref{topandwell_2d}--\ref{atopandwell_2d}.

%
% mfpts to the top of the barrier and to the other minima (2d plot)
%

\subsection{ MFPT-ratio no longer obeying normal behavior}

\begin{figure}[!ht]
\includegraphics[angle=0, width=8.0cm, height=5cm]{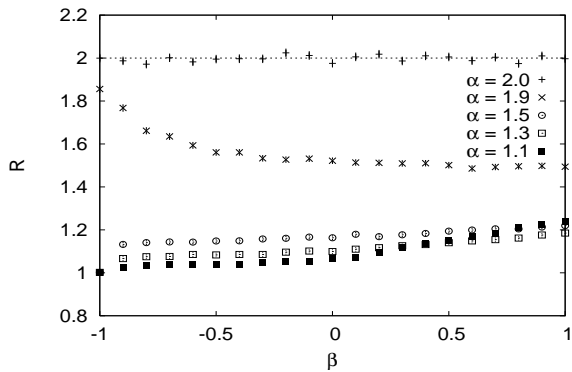}
\caption{Ratio $R=\MFPTw/\MFPTb$ of MFPTs to the neighboring
potential well ($\MFPTw$) and to the barrier top ($\MFPTb$). The
simulation parameters are: $dt=10^{-5}$, $N=2\times 10^4$,
$\sigma=\sqrt{2}$ and $N_b=2\times 10^3$.} \label{factor_2d}
\end{figure}

Next, we investigate the behavior of the value for the ratio
 $R=\MFPTw/\MFPTb$ of the $\MFPTw$ between the case with well-to well and well-to-bottom.
Note that from the exact expression in Eq.~(\ref{mfpttheory}), for
a symmetric potential $V(x)$ with normal Gaussian white noise
fluctuations (i.e. for $\alpha=2$ and an arbitrary value of the
skewness parameter $\beta$) this ratio yields the commonly known
factor of $R=2$. Put differently, in this case the random walker
needs twice the time to reach the other potential minimum than to
reach to the top of the potential barrier. In contrast, with a
decreasing value of $\alpha$ the ratio $R$ now {\it distinctly}
deviates from $2$, being always smaller than for a Gaussian
diffusion case. Our numerical results are depicted in
Fig.~\ref{factor_2d}. This deviation can be understood by noticing
that for $\alpha<2$, the stochastic escape trajectories of the
random walks in the double well potential become discontinuous,
meaning that the continuous movement of a particle becomes
interrupted with long jumps. These occasional jumps are more
probable to occur for small stability index $\alpha$ and explain
the observable statistics of the first passage times, implying in
turn that $R<2$. In particular, an increasing probability of long
jumps over the barrier in the overall statistics of passages tends
to equalize $\MFPTw$ with $\MFPTb$, yielding $R$ values
approaching $1$. The effect of the skewness parameter $\beta$
which is responsible for the asymmetric stochastic dynamics can be
interpreted as an additional contribution to the directionality of
random impact pulses which push the L\'evy-Brownian particle
within the potential well. For negative (positive) $\beta\approx
\pm1$ with $\alpha<1$, the driving L\'evy white noise becomes a
one-sided L\'evy process with strictly negative (positive)
increments which stabilize trajectories around the starting
position $x=\pm \xw$.

%
% ratio mfpt(well)/mfpt(top) (2d plot)
%

%%
%% conclusion
%%
\section{Conclusions}

In this paper we have studied the survival probabilities and the
mean values of the mean first passage times for escape from a
symmetric double well potential when the overdamped dynamics is
driven by general L\'evy white noise. The statistics of escaping
trajectories is investigated by a numerical analysis of a L\'evy
noise driven Langevin equation with properly implemented boundary
conditions. In contrast to former studies
\cite{chechkin2005,imkeller} aimed to understand statistics of
barrier crossings with symmetric L\'evy flights or exploring the
first passage problem for one-sided L\'evy motions \cite{eliazar},
our work provides the first analysis of the Kramers problem with an
arbitrary set of $\alpha, \beta$-parameters which characterize the
white stable noise-source entering the Langevin dynamics. Following
the Markovian character of the stochastic dynamics, -- at
sufficiently high barriers --, the time dependence of the survival
probabilities within the potential well assume an exponential law.
As can be expected, distinct deviations from the exponential
behavior become detectable, however, for low barrier heights when,
similarly to a normal Gaussian diffusion case, many recrossing of
the barrier are possible.

An asymmetry of the L\'evy noise perturbing the dynamics is shown
to either enhance or also suppress the escape events and
corroborates the intuitive influence of the $\beta$ parameter which
yields an additional, biasing contribution to the particles' motion
due to the directionality of stochastic impact pulses. Moreover, the
rate of escape is shown to exhibit a non-monotonic behavior as a
function of the stability index $\alpha$, with a discontinuity
occurring at the value $\alpha=1$.

In clear contrast to a normal diffusion behavior as typified by
systems driven by white Gaussian noise, for which the random walker
requires twice the time to reach the other potential minimum as
compared to the mean time to reach to the (absorbing) top of the
barrier, L\'evy white noise with $\alpha<2$ now causes a decrease in
this ratio of the mean first passage times: The ratio
$R=\MFPTw/\MFPTb$ lies consistently below the ``normal'' value of
$R=2$. This remarkable result holds true for any chosen value of the
skewness parameter $\beta$.

In summary, our numerical considerations demonstrate the richness of
the bi-stable kinetics resulting from driven $\alpha$-stable white
noises. The observed features are rooted in the non-local jump
lengths taken from a distribution which exhibit fat tails which in
turn rule the random passages between the attracting states of a
double well potential.

%%
%% acknowledgements
%%
\begin{acknowledgments}
The Authors acknowledge the financial support from the Polish
State Committee for Scientific Research (KBN) through the grants
1P03B06626 (2004--2005) and 2P03B08225 (2003--2006), Marie Curie
TOK COCOS grant (6th EU Framework Programme under contract: MTKD-CT-2004-517186)
 and the ESF funds (E.G.N. and P.H.) via the
STOCHDYN program. Additionally, B.D. acknowledges the financial
support from the Foundation for Polish Science through the domestic
grant for young scientists (2005). Computer simulations have been
performed at the Academic Computer Center CYFRONET AGH, Krak\'ow.
P.H. also acknowledges the support by the German research foundation
(DFG) via grant Ha 1517/26-1, 2 (``Single molecules'').
\end{acknowledgments}


\begin{thebibliography}{99}
\bibitem{foot}
The result follows a general observation for separable continuous
time random walks (CTRW) which lead to normal stochastic diffusion
for finite mean waiting time for a jump to occur and finite step
variance \cite{zolotarev}.
\end{thebibliography}
\end{document}